\documentclass[10pt]{article}

\usepackage{float,xspace,amsmath,amssymb,color,graphicx}

\def\band{\textrm{\sc and}\xspace}
\def\bnot{\textrm{\sc not}\xspace}

\def\nil{\textrm{\sc null}\xspace}

\newcommand{\nstr}[1]{{\overline{#1}}}

\newcommand{\uassign}{\leftarrow}
\newcommand{\uif}{{\bf if}\xspace}
\newcommand{\uthen}{{\bf then}\xspace}
\newcommand{\uelse}{{\bf else}\xspace}
\newcommand{\uwhile}{{\bf while}\xspace}

\newcommand{\ucontinue}{{\bf continue}\xspace}
\newcommand{\ufor}{{\bf for}\xspace}
\newcommand{\uto}{{\bf to}\xspace}

\newcommand{\udo}{{\bf do}\xspace}
\newcommand{\ureturn}{{\bf return}\xspace}

\newcounter{lineno}

\newcommand{\uln}{\stepcounter{lineno}\thelineno\>}

\newcommand{\utab}{4ex}

\newenvironment{code}{
\setcounter{lineno}{0}
\begin{tabbing}
\utab\=\utab\=\utab\=\utab\=\utab\=\utab\=\utab\=%
\utab\=\utab\=\utab\=\utab\=\utab\=\utab\= \kill
}
{
\end{tabbing}
\vspace{-2ex}
}

\floatstyle{ruled}
\newfloat{algorithm}{tbp}{flt}
\floatname{algorithm}{\small\bf Alg.}

\newtheorem{problem}{Problem}

\begin{document}

\title{On building minimal automaton for subset matching queries}

\author{Kimmo Fredriksson\\
School of Computing, University of Eastern Finland,\\
P.O. Box 1627, 70211 Kuopio, Finland\\
{\tt kimmo.fredriksson@uef.fi}
}

\date{}

\maketitle

\thispagestyle{empty}

\begin{abstract}
\noindent
We address the problem of building an index for a set $D$ of $n$
strings, where each string location is a subset of some finite integer
alphabet of size $\sigma$, so that we can answer efficiently if a
given simple query string (where each string location is a single
symbol) $p$ occurs in the set. That is, we need to efficiently find a
string $d \in D$ such that $p[i] \in d[i]$ for every $i$. We show how
to build such index in $O(n^{\log_{\sigma/\Delta}(\sigma)}\log(n))$
average time, where $\Delta$ is the average size of the subsets.  Our
methods have applications e.g.\ in computational biology (haplotype
inference) and music information retrieval.
\end{abstract}

\noindent
{\bf Keywords:}
algorithms;
approximate string matching;
subset matching;
finite-state automaton minimization

\section{Introduction}

Let $\Sigma = \{0, \ldots, \sigma-1\}$ be an ordered integer alphabet.
We are given a set $D = \{d_0, \ldots, d_{n-1}\}$ of strings, called a
{\em dictionary}.  Each location $j$ of the string $d_i$ is a subset
of $\Sigma$, i.e.\ $d_i[j] \subseteq \Sigma$ for every $0 \leq i \leq
n-1$ and $0 \leq j \leq |d_i|-1$. A string $p$ is called {\em simple}
if its each location is a single symbol from $\Sigma$, i.e.\ $p[j] \in
\Sigma$. The simple {\em query} string $p$ {\em matches} the
dictionary string $d_i \in D$ iff $p[j] \in d_i[j]$ for $0 \leq j \leq
|p|-1$ and $|p| = |d_i|$.
We consider the following two problems:
\begin{problem}
Decide if $p$ matches any string in $D$.
\end{problem} 
\begin{problem}
Retrieve the set $L = \{j_1, \ldots, j_r\}$ such that $p$ matches
$d_{j_i}$ for all $j_i \in L$.
\end{problem}
In particular, we set out to {\em efficiently build a small {\em
index} for $D$ such that both problems can be solved in $O(|p|)$
time.}

Efficient solution of these problems have applications in
computational biology, in matching DNA ($\sigma = 4$) or protein
($\sigma=20$) strings, or in haplotype inference ($\sigma=2$)
\cite{DBLP:conf/cpm/Gusfield03,DBLP:conf/spire/LandauTW07}.
Finally, note that if $|d_i[j]|$ is either $1$ or $\sigma$ for all $i,
j$, then we have a special case called {\em wild-card matching}
\cite{coleetal}.
Another special case is $\delta$-{\em matching} (see e.g.\
\cite{DBLP:journals/ijcm/CambouropoulosCIMP02}), where we have $d_i[j]
= \{c_{i,j}-\delta, \ldots, c_{i,j}+\delta\}$ where $c_{i,j} \in
\Sigma$, and $\delta < \sigma$.
These variants have applications in indexing natural language words
and in music information retrieval.

\subsection{Related work}

Assume that the longest string in $D$ has length $m$ and that for
every $d_i \in D$ there are at most $k$ locations where $|d_i[j]| >
1$. The immediate trivial solution to our problem would then be as
follows. First generate all the simple strings of length $m$ that
match a string in $D$. Call the set of these strings $D'$.  The size
of $D'$ is upper bounded by $O(n\sigma^k)$.
The problem is now transformed to exact matching, so we can insert all
strings in $D'$ to some data structure that can answer whether a given
simple query string matches a string in the data structure {\em
exactly}. One such data structure is a path compressed {\em trie}
\cite{Fre60} (cf.\ Sec.~\ref{sec:alg}).  This can be na\"ively built
in $O(m|D'|) = O(mn\sigma^k)$ time and space.  The queries can be
answered in $O(|p|)$ time.

This is also the approach in \cite{DBLP:conf/spire/LandauTW07}. They
give two non-trivial algorithms to construct the (path compressed)
trie faster, namely in $O(nm+\sigma^k n \log(\min\{n,m\}))$ and
$O(nm+\sigma^kn + \sigma^{k/2}n \log(\min\{n,m\}))$ time, yielding
query times of $O(|p|)$ and
$O(|p|\log\log(\sigma)+\min\{|p|,\log(\sigma^kn)\}\log\log(\sigma^kn))$
respectively (the latter method in fact uses two tries).

The techniques in \cite{coleetal} can be adapted
\cite{DBLP:conf/spire/LandauTW07} to solve the problem with
$O(nm\log(nm) + n\log^k (n/k!))$ preprocessing time, and
$O(m+\log^k(n)\log\log(n))$ query time.

\subsection{Our contributions}

Inspired by \cite{DBLP:conf/spire/LandauTW07}, we also take the
approach of computing the trie for $D'$ as a starting point. However,
instead of a trie, we {\em directly} build a {\em pseudo-minimal} (cf.\ 
Sec.~\ref{sec:pmDFA}) deterministic finite-state automaton (DFA)
corresponding to the set $D'$; 
i.e.\ our method does not
explicitly generate the set $D'$.
The resulting automaton can be used to solve Problems~1 and 2 in
$O(|p|)$ time. This automaton can be easily and efficiently
minimized (again, cf.\ Sec.~\ref{sec:pmDFA}),
so that the Problem~1 can still be solved in $O(|p|)$ time.  We also
propose a form of path compression that can further save space and
speed up the construction.  We show that our construction works in
$O(n^{\log_{\sigma/\Delta}(\sigma)}\log(n))$.  average time, where
$\Delta = \text{avg} \; |d_i[j]|$.

As shown experimentally, our algorithm can be orders of magnitude
faster in construction time than the related na\"ive approach of first
building a trie for $D'$, and then converting it to the minimal DFA,
or directly building the minimal DFA from $D'$.
The pseudo-minimal automaton is more efficient to construct than the
true minimal automaton, and is in practice only slightly larger.

\section{The algorithm} \label{sec:alg}

Let us define a DFA as $M(Q,\Sigma,\delta,q,F)$, where $Q$ is the set
of states, $q$ is the initial state, $F \subseteq Q$ is the set of
accepting states and $\delta \in Q \times \Sigma \rightarrow Q$ is the
transition function.  For convenience we also define $\delta^*(q,aw) =
\delta^*(\delta(q,a),w)$ for a string $w \in \Sigma^*$.

\subsection{Prelude}

Traditionally a {\em trie} \cite{Fre60} is described as being a rooted
tree storing a set of (simple) strings. Each node has at most $\sigma$
children, and the (directed) edges are labeled by the symbols in
$\Sigma$. In {\em path compressed} trie the unary paths are compacted
to single edges, labeled by strings consisting of the concatenation of
the symbols in the original path. In both cases, a path from the root
to any node $u$ spells out a prefix of a subset of the strings stored
in the trie, and that subset is stored in the subtree rooted at $u$. The
trie can be seen as a DFA in an obvious way; the root node
corresponding to the state $q$, and the labeled edges corresponding to
$\delta$.

We extend the DFA so that for the nodes $u \in F$ we attach a list
$L$, storing the corresponding string identifiers. More formally, we
define
\begin{equation}
j_i \in L(u) ~ \Leftrightarrow ~ \nstr{u} \text{ matches } d_{j_i} \in D,
\label{eq:L}
\end{equation}
where $\nstr{u}$ denotes the string spelled by the path from $q$ to
$u$, i.e.\ $\nstr{u} = (w\;|\;\delta^*(q,w) = u)$. Thus by generating
all the strings $D'$ that match a string in $D$ and building a DFA for
$D'$, Problems~1 and 2 can be solved in $O(|p|)$ time.

One of the problems of this approach is that $|D'|$ can be large. A
way to alleviate this is to minimize the DFA. There exists a large
number of algorithms for this task \cite{DBLP:conf/wia/Daciuk02}.
Some of these can build the automaton incrementally, inserting one
string at a time while maintaining the automaton in minimal state
(e.g.\ \cite{DBLP:journals/coling/DaciukMWW00}).

This can still be unnecessarily slow. Moreover, the result does not
allow proper mapping between the states and the lists $L$.  E.g.\ 
if all the strings in $D$ are of equal length, the resulting
minimal DFA would have only one accepting state. However, this
automaton can still be used to solve Problem~1. Another solution is to
construct a {\em pseudo-minimal} DFA \cite{Maurel2000129,
DBLP:conf/wia/DaciukMS05} still allowing mapping states or transitions
to strings. We take a similar approach, although our definition of
pseudo-minimal is somewhat different.

\subsection{Pseudo-minimal DFA} \label{sec:pmDFA}

We now present an algorithm that directly (i.e.\ our algorithm never
deletes a state) constructs pseudo-minimal DFA from $D$, without using
a trie-like DFA as an intermediate step, or explicitly generating the
set $D'$.
Nevertheless, we first describe a particular (direct) way to build a
trie-DFA, and then define a certain equivalence relation for the trie
states, and show how we can during the construction avoid creating new
states by identifying an equivalent state already present.

The algorithm can proceed recursively in either a depth-first or a
breadth-first manner, with minor differences. We describe and give
pseudo code for the breadth-first variant: the construction begins by
inserting the starting state (root node) into {\em queue} of states; 
at each stage a state is dequeued and its children are computed and
enqueued. The algorithm terminates when the queue becomes empty.
As described above, each state $u$ will have an associated list
$L(u)$, ($L(u) = \emptyset$, if $u \not\in F$).  We will denote the
partially computed list as $L'(u)$ ($L'(u) \neq \emptyset$). 
The following invariants are
maintained: (a) when all the children (if any) of $u$ are enqueued,
the state $u$ is fully computed and Eq.~(\ref{eq:L}) is satisfied
(post-condition); (b) when a state $u$ is enqueued, then the list
$L'(u)$ satisfies Eq.~(\ref{eq:Lp}) below (pre-condition):
\begin{equation}
j_i \in L'(u) ~ \Leftrightarrow ~ \nstr{u} \text{ matches } d_{j_i}[0 \ldots |\nstr{u}|-1] ~ | ~ d_{j_i} \in D.
\label{eq:Lp}
\end{equation}
I.e.\ $j_i \in L'(u)$ iff $\nstr{u}$ matches a prefix of $d_{j_i}$
(note that $|\nstr{u}| = depth(u)$, if the paths are not
compressed). Thus the algorithm initializes
\begin{equation}
L'(q) = \{0, \ldots, n-1\}
\end{equation}
and enqueues $q$. At each iteration, one state $u$ is dequeued, its
``children'' are initialized according to the pre-condition, and
enqueued, and the post-condition for $u$ is computed. Given the list
$L'(u)$ and $\forall c \in \Sigma$, we define
\begin{equation}
L'(v) = \{ j_i \;|\; j_i \in L'(u) ~ \band ~ c \in d_{j_i}[|\nstr{u}|] \}.
\end{equation}
If $|L'(v)| > 0$, then a transition $\delta(u,c) = v$ is added, and
$v$ enqueued.  Note that $j_i$ is put into $|d_{j_i}[|\nstr{u}|]|$
lists.  The list $L(u)$ is then computed as
\begin{equation}
L(u) = \{ j_i \;|\; j_i \in L'(u) ~ \band ~ |\nstr{u}| = |d_{j_i}| \}.
\end{equation}
That is, we keep only the strings that end in $u$, and $u$ becomes an
accepting state iff $|L(u)| > 0$.  All the $\sigma$ lists $L'(v)$ and
the list $L(u)$ can be computed with a single pass over the the list
$L'(u)$. Alg.~\ref{alg:partition} gives the pseudo code.

This is repeated until the queue becomes empty. Note that this
computes exactly the same trie as one would get by first generating
$D'$ and then inserting the strings one at a time. However, our
bulk-insertion method is more easily improved.

We define the following relation between the states $u$ and $v$:
\begin{equation}
u \equiv_p v: ~ L'(u) = L'(v) ~ \band ~ |\nstr{u}| = |\nstr{v}|,
\end{equation}
which is clearly reflexive, symmetric, and transitive, i.e.\ an equivalence relation.
The following is easy to notice:
\begin{equation}
u \equiv_p v \Rightarrow {\cal L}(u) = {\cal L}(v),
\end{equation}
where the {\em language} of $u$ is
\begin{equation}
{\cal L}(u) = \{w \in \Sigma^* \;|\; \delta^*(u,w) \in F\}. \label{eq:lang}
\end{equation}

Hence we will partition the states into equivalence clas\-ses, so that
in the final DFA all states belong to a different class.
Note that this does not result in a minimal DFA; i.e.\ we have that
${\cal L}(u) = {\cal L}(v) \nRightarrow u \equiv_p v$, while the
implication would be required for a true minimal automaton.
Note that by the definition we can still properly associate states
with the lists $L'$ and $L$. So we can call the result {\em
pseudo-minimal} DFA as in
\cite{Maurel2000129,DBLP:conf/wia/DaciukMS05}, even when our
definition should not be confused with the definition given in these
papers

We need to maintain sets of pairs $(L',u)$, where $L'$ is a key that
is used to insert and search the state $u$, a representative of its
equivalence class.  The algorithm is now immediate: whenever we have
computed a list $L'(v)$, we search if it is present in a set
$S(depth(v))$; if so, $v$ can be replaced by the corresponding node
$u$. In this case, $v$ is not enqueued,
as an equivalent state $u$ is in the queue already.
If $L'(v)$ is not present, we
insert $(L'(v),v)$ to $S(depth(v))$, and enqueue $v$.
Alg.~\ref{alg:build} gives the complete pseudo code, keeping the
automaton in its pseudo-minimal state throughout the construction.

\begin{algorithm}[!t]
\begin{code}
\uln \>	$L \uassign \emptyset$ \\
\uln \>	\ufor $c \uassign 0$ \uto $\sigma-1$ \udo $P[c] \uassign \emptyset$ \\
\uln \>	\ufor $i \uassign 0$ \uto $|L'|-1$ \udo \\
\uln \>	\>	$k \uassign L'[i]$ \\
\uln \>	\>	\uif $|d_{k}| \leq depth$ \uthen \\ 
\uln \>	\>	\>	$L \uassign L \cup \{k\}$ \\
\uln \>	\>	\uelse \\
\uln \>	\>	\>	\ufor $\forall c \in d_{k}[depth]$ \udo 
$P[c] \uassign P[c] \cup \{k\}$ \\
\uln \>	\ureturn $(L,P)$
\end{code}
\caption{Partition($D, L', depth$).}
\label{alg:partition}
\end{algorithm}

\begin{algorithm}[!t]
\begin{code}
\uln \>	$q \uassign \text{NewState}()$ \\
\uln \>	$L'(q) \uassign \{0,\ldots,|D|-1\}$ \\
\uln \>	$\text{Enqueue} (q)$ \\
\uln \>	\uwhile \bnot $\text{QueueEmpty}()$ \udo \\
\uln \>	\>	$u \uassign \text{Dequeue}()$ \\
\uln \> \>	$(L(u),P) \uassign \text{Partition} (D, L'(u), depth(u))$ \\
\uln \>	\>	\uif $|L(u)| > 0$ \uthen $F \uassign F \cup \{u\}$ \\
\uln \> \>	\ufor $c \uassign 0$ \uto $\sigma-1$ \udo \\
\uln \> \>	\>	\uif $|P[c]| = 0$ \uthen \ucontinue \\
\uln \>	\>	\>	$v \uassign \text{Search} (S [depth(u)], P[c])$ \\
\uln \>	\>	\>	\uif $v = \nil$ \uthen \\
\uln \>	\>	\>	\>	$v \uassign \text{NewNode}()$ \\
\uln \>	\>	\>	\>	$L'(v) \uassign P[c]$ \\
\uln \>	\>	\>	\>	$\text{Insert} (S[depth(u)], (L'(v), v))$ \\
\uln \>	\>	\>	\>	$\text{Enqueue}(v)$ \\
\uln \>	\>	\>	$\delta(u,c) \uassign v$ \\
\uln \>	\ureturn $q$
\end{code}
\caption{BuildDFA($D$).}
\label{alg:build}
\end{algorithm}

\subsection{Using subsets for unary paths} \label{sec:cutoff}

For a moment consider a plain trie with a path compression. In this
case the trie has $\Theta(|D'|)$ nodes (states), independent of the
pattern lengths (without path compression, this is multiplied by
$O(m)$). While this may save space in many cases, this is not always
so. Consider e.g.\ the unrealistically pathological case, where $D$
contains only one string of length $m$, namely $\Sigma^m$. This means
that all $\sigma^m$ possible strings are present in $D'$, and no path
compression can take place, as there simply are no unary paths (the
minimal and pseudo-minimal DFAs would both have only $m+1$ states). We
propose a slightly different, but much more effective, path
compression.

Consider now a string in $D$, and in particular that the string
positions can be any subsets of $\Sigma$ (not necessarily just single
symbols).  Assume that $d_i[depth(u)] = d_j[depth(u)]$, for some $u$
and $\forall i,j \in L'(u)$. This means that there is no need to
branch, since all the subsets are the same, and no symbol in $\Sigma$
can differentiate between any $d_i$, $d_j$. Hence we could add a
transition from $u$ to (some) $v$ using the {\em subset}
$d_i[depth(u)]$ as a label. 
This does not pose any problems, as (when used in
recognition) we can still test in $O(1)$ time if $p[depth(u)] \in
d_i[depth(u)]$.  (Note that our pseudo-minimization algorithm
effectively already handles this, i.e.\ under the above condition,
$\delta(u,c) = v$ for $\forall c \in d_i[depth(u)]$.)

More generally, given a node $u$, and
\begin{equation}
\forall i,j \in L'(u): d_i[k] = d_j[k] \;|\; depth(u) \leq k < h,
\end{equation}
then $d_i[depth(u) \ldots h-1]$ can be used as a string label in a
compressed unary path.

The easiest way to utilize this is to use it only for unary paths to
the leaves when $|L'(u)|=1$.  This is effectively achieved simply by
replacing the line 15 in Alg.~\ref{alg:build} by ``\uif $|L'(v)| > 1$
\uthen Enqueue$(v)$''. It would be relatively easy to use the path 
compression in any unary path, but as show in Sec.~\ref{sec:exp} this 
simple method can give huge savings in both time and space.

\subsection{Analysis}

Let us now consider the running time of Alg.~\ref{alg:build}, with
(our) path compression on leaves. We assume that the subsets $d_i[j]$
have average size $\Delta$, and that they are are randomly,
uniformly and independently generated. 
At first we assume that there is a non-zero probability that two
random subsets do not intersect (e.g.\ $\Delta \leq \sigma/2$).

The partition of $L'(u)$ can be implemented to take $O(|L'(u)|\Delta)$
time. Each of the $\sigma$ resulting new sets have average size
$O(|L'(u)|\Delta/\sigma)$, as for a random $c \in \Sigma$ the
probability that $c \in d_i[j]$ is $\Delta/\sigma$. These sets are
searched from $S$, and possibly inserted (if not found). The size of
$S$ is upper bounded by $O(|Q|)$, the number of states in the
resulting automaton. Hence insert/search can be implemented in
$O(\log(|Q|)+|L'(u)|\Delta/\sigma)$ worst case time with a number of
radix-tree techniques, see e.g.~\cite{patricia,balancedTST}. Therefore
the total time per node is $O(\sigma(\log(|Q|)+|L'(u)|\Delta/\sigma) +
|L'(u)|\Delta)$, i.e.\ $O(\sigma\log(|Q|)+|L'(u)|\Delta)$, which is
$O(\log(|Q|)+|L'(u)|)$, assuming $\sigma = O(1)$.

For a moment assume that we are building a plain trie, without path
compression.  Recall that by definition the length of the list
$L'(root)$ is exactly $n$.  
As described above, the length\footnote{In the
``worst case'' there is only one ``new'' set, being exactly the same as its
parent, $L'(u)$; but in this case the corresponding node would not branch,
so the complexity would only improve.} of each of
the $\sigma$ lists for the children of node $u$ is
$O(|L'(u)|\Delta/\sigma)$, so the lengths of the lists $L'(u)$
decrease exponentially when the depth of $u$ (i.e.\ $|\nstr{u}|$)
increase, as $|L'(u)| = O((\Delta/\sigma)^{|\nstr{u}|}n)$. 
Hence
$|L'(u)| = O(1)$ when $\alpha = |\nstr{u}| \geq 
\log_{\sigma/\Delta}(n)$. The total number of states up to
this depth is $|Q| = \sum_i^\alpha \sigma^i = O(\sigma^\alpha) =
O(n^{\log_{\sigma/\Delta}(\sigma)})$, that is, all the states have all
the $\sigma$ possible branches up to depth $\alpha$.
As there are $\sigma^i$ nodes at depth $i$, the total length of {\em
all} the lists at a depth $i$ is on average $O(({\Delta}/{\sigma})^i n
\; \sigma^i) = O(\Delta^i n)$.  Thus the total length of all the lists
up to depth $\alpha$ is $\ell = n \sum_i^\alpha \Delta^i = O(n
\Delta^\alpha) = O(n^{\log_{\sigma/\Delta}(\Delta)+1}) =
O(n^{\log_{\sigma/\Delta}(\sigma)})$.

Assume now (pessimistically) that path compression and
pseudo-minimization take place only after depth $\alpha$. After this
depth, the lists have length $k=O(1)$, (and will continue to shrink
until $k=1$).There are only
$\binom{n}{k}=O(n^k/k!)$ different lists of length $k$, but at the
same time there are $O(n^{\log_{\sigma/\Delta}(\sigma)})$ states (with
associated lists), so by the pigeonhole principle many of the states
must be equivalent, and are combined into a single state. 
However, due to path compression, the process terminates for any state
having $k=1$.
Hence the number of states per level starts to decrease
exponentially\footnote{Note that without combining the equivalent
states or the path compression, after depth $\alpha$ the number of
states would continue to {\em increase} exponentially, resulting in a
full trie.} after depth $\alpha$. That is, the total number of states
is bounded by two geometric series, both having the largest 
term at depth $\alpha$, where the automaton is in its ``widest'', 
i.e.\ the total number of states is asymptotically upper bounded by
$O(n^{\log_{\sigma/\Delta}(\sigma)})$.

Summing up, the total time is on average 
\begin{equation}
O(|Q|\log(|Q|)+\ell) =
O(n^{\log_{\sigma/\Delta}(\sigma)}\log(n)),
\end{equation}
again assuming $\sigma = O(1)$. 

So far we have assumed that there is a non-zero probability that two
random subsets do not intersect. Consider now the (rather uninteresting)
case where the subset sizes are {\em always} $\Delta>\sigma/2$ (not 
just on average). 
At first, the process goes as before, the number of states increasing
exponentially, and the list lengths $|L'(u)|$ decreasing exponentially.
However, assume now, for simplicity, that $L'(u) = \{i,j\}$ for some state
$u$. Due to $\Delta>\sigma/2$, the subsets $d_i[h]$ and $d_j[h]$ must
intersect (where $h = |\nstr{u}|$).
Thus the alphabet $\Sigma$ is effectively partitioned into four 
disjoint sets:
$A = d_i[h] \setminus d_j[h]$; 
$B = d_j[h] \setminus d_i[h]$; 
$C = d_i[h] \cap d_j[h]$; 
$D = \Sigma \setminus (d_j[h] \cup d_i[h])$. Group $D$ does generate
any branches for $u$. Symbols from $A$, $B$ and $C$ generate branches,
but these are combined (group-wise) by the minimization, resulting in at
most one new state per group, call it $v$. For $A$ (similarly for $B$),
$|L'(v)|=1$, and due to path compression, $v$ will have no descendants. 
The interesting case is $C$. Note that $C$ cannot be empty, so
$L'(v) = L'(u)$, and hence the process repeats for $v$. In other words,
the process does not terminate until $h = |d_i|$. 

The situation is similar when $|L'(u)| > 2$. Note that after depth $\alpha$
we have $|L'(u)| = O(\Delta)$ in any case, and because of the 
pseudo-minimization, there can be only 
$\binom{n}{\Delta} = O(n^\Delta/\Delta!)$
different states with lists of length $O(\Delta)$. Thus 
in general the ``breadth'' of the
automaton will stay approximately the same after depth $\alpha$, and the 
total time is upper bounded by
$O((n^{\log_{\sigma/\Delta}(\sigma)}+\min\{n^{\log_{\sigma/\Delta}(\sigma)}, n^\Delta\}\;m)\log(n))$,
where $m$ is the length of the strings in $D$.

Finally, as the number of subsets of $n$ items is at most $2^n$, the 
trivial upper bound for the worst case size of our data structure is
$O(m2^n)$. This should be contrasted with the $O(\sigma^m)$ bound of
\cite{DBLP:conf/spire/LandauTW07}.

\section{Experiments and final remarks} \label{sec:exp}

We have implemented the algorithms in C, and ran the experiments on
3.0GHz Intel Core2 with 2GB RAM, 4MB L2 cache, running GNU/Linux
2.6.23.

The implemented algorithms are: Pseudo-minimal DFA (PM DFA), as in
Alg.~\ref{alg:build}; 
minimal DFA (M DFA);
PM DFA with path compression (PM DFA PC) on leaves, as detailed in
Sec.~\ref{sec:cutoff}; plain trie; and trie with path compression on
leaves (Trie PC), as in PM DFA PC.  Some results for the Tries are not
included, as they could not fit into the available RAM. M DFA was 
computed from PM DFA, as computing it from
$D'$ or the corresponding trie would have been totally intractable
in most cases.
We implemented the set $S$ in Alg.~\ref{alg:build} with Patricia tries
\cite{patricia}.

We have not implemented the methods in
\cite{DBLP:conf/spire/LandauTW07}, but we show that the lower bound
($|D'|$) for the size of their data structure can be several orders of
magnitude larger than our empirical sizes. In fact, we can build
reasonably small data structures for problem instances that are
completely intractable with their methods. 

Table~\ref{table:r} gives the results for some randomly generated
instances. We used parameters $(m,n,\sigma,(\Delta_l,\Delta_h),f)$,
where $m$ is the length of the strings (all $n$ of equal length);
$(\Delta_l,\Delta_h),f$ denotes that in probability $f$ any string
location contains a randomly selected subset of $\Sigma$, where the
size of the subset is randomly selected between $\Delta_l \ldots
\Delta_h$; otherwise (with probability $1-f$) the string location is a
single random symbol from $\Sigma$.

\begin{table}[!t]
\caption{Experimental results for data generated using parameters  $(m,|D|,\sigma,(\Delta_l,\Delta_h),f)$. Times are given in seconds.
$|Q|$ is the number of generated states, and $|D'|$ is the number of 
different strings matching a string in $D$.}
\label{table:r}
\centerline
{
\begin{tabular}{|c|c|r|}
\hline
\multicolumn{3}{|c|}{$(32, 10000, 2, (2,2), 0.2)$,
$|D'| = 3,418,449$} \\
\hline
Method		& time (s) & $|Q|$ \\
\hline
\hline
PM DFA		& 0.828	& 476,365  \\
\hline
M DFA		& 1.20	& 385,255 \\
\hline
PM DFA PC	& 0.820	& 379,948  \\
\hline
Trie		& 6.71	& 18,767,894 \\
\hline
Trie PC		& 0.326	& 948,493  \\
\hline
\multicolumn{3}{c}{} \\
\hline
\multicolumn{3}{|c|}{$(32, 10000, 4, (2,4), 0.3)$,
$|D'| = 40,755,624,312$} \\
\hline
\hline
PM DFA		& 1.42	& 680,906  \\
\hline
M DFA		& 2.39	& 635,795 \\
\hline
PM DFA PC	& 1.40	& 499,212  \\
\hline
Trie PC		& 1.19	& 4,203,673 \\
\hline
\multicolumn{3}{c}{} \\
\hline
\multicolumn{3}{|c|}{$(16, 10000, 20, (2,6), 0.75)$,
$|D'| = 1,830,872,526,457$}\\
\hline
\hline
PM DFA		& 6.50	& 1,335,251 \\
\hline
M DFA		& 18.2	& 1,320,126 \\
\hline
PM DFA PC	& 6.21	& 1,276,985 \\
\hline
Trie PC		& 6.29	& 22,431,630 \\
\hline
\multicolumn{3}{c}{} \\
\hline
\multicolumn{3}{|c|}{$(16, 100000, 32, (2,32), 0.01)$,
$|D'| = 1,033,039$} \\
\hline
\hline
PM DFA		& 6.28	& 1,331,241  \\
\hline
M DFA		& 12.0	& 964,847  \\
\hline
PM DFA PC	& 0.486	& 149,998  \\
\hline
Trie		& 15.5	& 6,981,214  \\
\hline
Trie PC		& 0.236	& 235,565  \\
\hline
\multicolumn{3}{c}{} \\
\hline
\multicolumn{3}{|c|}{$(16, 1000, 32, (32,32), 0.25)$,
$|D'| \approx 1,190 \times 10^{15}$} \\
\hline
\hline
PM DFA		& 0.954	& 118,474  \\
\hline
M DFA		& 7.29	& 115,797  \\
\hline
PM DFA PC	& 0.907	& 110,340  \\
\hline
\end{tabular}
}
\end{table}

\begin{figure}[!ht]
\begin{center}
\includegraphics[width=0.98\textwidth]{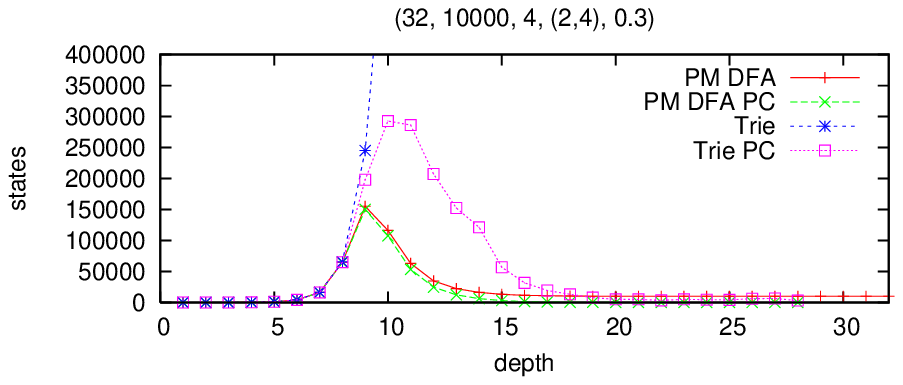}
\includegraphics[width=0.98\textwidth]{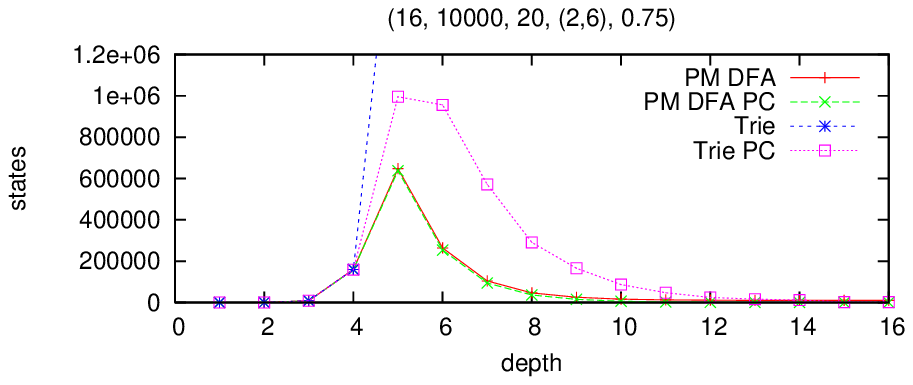}
\includegraphics[width=0.98\textwidth]{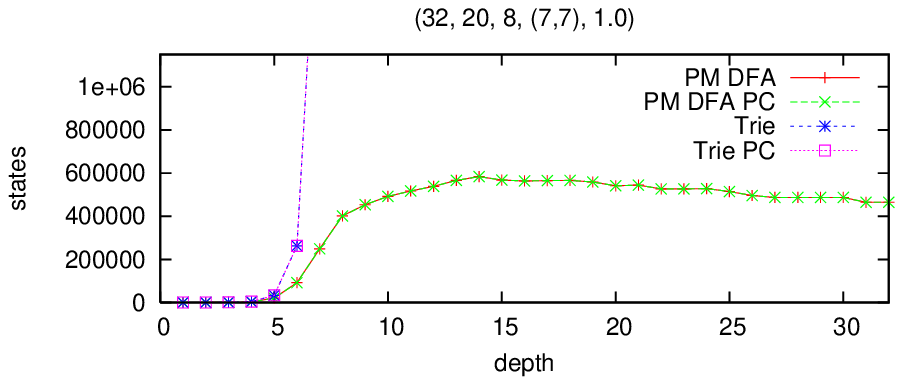}
\end{center}
\caption{The total number of states generated for each depth of the 
automaton / trie during
the construction. Above: $\alpha = \log_{\sigma/\Delta}(n) \approx 10.04$; 
middle: $\alpha \approx 5.07$. Below: subset sizes always $> \sigma/2$.}
\label{fig:r}
\end{figure}

We report the number of states generated by the different methods, as
well as the time in seconds, for some illustrative cases.
As shown, the number of states generated is significantly smaller than
$|D'|$ in all cases, sometimes the difference being many orders of
magnitude. PM DFA is usually only slightly larger than the
true minimal DFA, while using path compression with PM DFA is usually
{\em smaller} than M DFA. In some rare cases using path
compression with a plain Trie is very competitive. Fig.~\ref{fig:r}
shows the exponential increase (depth $\lessapprox\alpha$) and
decrease (depth $\gtrapprox \alpha$) of the number of states as a
function of the depth in the automaton / trie, and illustrates
the behaviour when all subset sizes are $> \sigma/2$.

Finally, we note that our methods have applications in on-line
dictionary string matching, e.g.\ in $\delta$-matching and
$(\delta,\gamma)$-matching. It turns out that we can solve both
problems in $O(|T| \log_{\sigma/\delta}(nm)/m)$ average time, which is
optimal for $\delta$-matching \cite{FMN06ijfcs}, for a dictionary of
$n$ patterns of length $m$, and a text of length $|T|$. We leave the
details for future work.

\end{document}